\documentclass[aps,prl,reprint,showpacs,superscriptaddress]{revtex4-1}

\usepackage{amsmath,amssymb,times,bm}
\usepackage{graphicx,color}
\usepackage[pdfborder={0 0 0},colorlinks=true,citecolor=blue,urlcolor=blue]{hyperref}
\graphicspath{{./fig/},{./}}

\begin{document}
\title{Majorana bound states and non-local spin correlations\\ in a quantum wire
on an unconventional superconductor}
\date{\today}
\author{Sho Nakosai}
\affiliation{Department of Applied Physics, University of Tokyo, Tokyo 113-8656, Japan}
\author{Jan Carl Budich}
\affiliation{Department of Physics, Stockholm University, Se-106 91 Stockholm, Sweden}
\author{Yukio Tanaka}
\affiliation{Department of Applied Physics, Nagoya University, Aichi
464-8603, Japan}
\author{Bj\"{o}rn Trauzettel}
\affiliation{Institute for Theoretical Physics and Astrophysics,
University of W\"{u}rzburg, 97074 W\"{u}rzburg, Germany}
\author{Naoto Nagaosa}
\affiliation{Department of Applied Physics, University of Tokyo, Tokyo 113-8656, Japan}
\affiliation{Correlated Electron Research Group (CERG), RIKEN Advanced
Science Institute (ASI), Wako 351-0198, Japan}
\affiliation{Cross-Correlated Materials Research Group (CMRG), RIKEN
Advanced Science Institute (ASI), Wako 351-0198, Japan}

\begin{abstract}
 We study theoretically the proximity effect of a one-dimensional
 metallic quantum wire (in the absence of spin-orbit interaction) lying
 on top of an unconventional superconductor. Three different material
 classes are considered as a substrate: (i) a chiral superconductor in
 class D with broken time-reversal symmetry; a class DIII superconductor
 (ii) with and (iii) without a nontrivial $\mathbb{Z}_2$
 number. Interestingly, we find degenerate zero energy Majorana bound
 states at both ends of the wire for all three cases. They are unstable
 against spin-orbit interaction in case (i) while they are topologically
 protected by time-reversal symmetry in cases (ii) and (iii). 
 Remarkably, we show that non-local spin correlations between the two
 ends of the wire can be simply controlled by a gate potential in our
 setup.
\end{abstract}
\maketitle

\textit{Introduction.--}
Proximity effects of superconductors and normal metals as well as
semiconductors have been a subject of continued interest due to the rich
physical phenomena of these hybrid systems. Especially, the topological
nature of the superconducting proximity effect is an issue of current
research activities~\cite{PhysRevLett.100.096407, arXiv:1112.1950,
JPSJ.81.011013, RepProgPhys.75.076501} stimulated by the possible
occurrence of Majorana fermions in solid state
systems~\cite{NuovoCim.14.171, NatPhys.5.614, Physics.3.24,
PhysRevB.61.10267, PhysRevLett.86.268, PhysRevLett.94.166802,
RevModPhys.80.1083}.
For example, the interface between an $s$-wave superconductor and the
surface state of a three-dimensional topological insulator (TI) is
predicted to offer a platform to realize Majorana
fermions~\cite{PhysRevLett.100.096407}.
The underlying idea here is holographic principle that entails the
reduction of electronic degrees of freedom, i.e., electron
fractionalization, at the surface of a topologically non-trivial bulk
state. Furthermore, the proximity effect of a TI to unconventional
superconductors has also been studied
theoretically~\cite{PhysRevLett.104.067001}.
Another (highly interesting) proposal is to use semiconductors with
Rashba spin-orbit interaction in combination with $s$-wave
superconductors for this purpose. In the presence of a sufficiently
strong magnetic field, an inverted gap opens at the $\Gamma$-point in
this system and topological superconductivity as well as Majorana
fermions may appear when the Fermi energy lies inside the
gap~\cite{PhysRevLett.103.020401, PhysRevLett.104.040502,
PhysRevB.81.125318, PhysRevLett.105.177002}.
A recent experiment observed the zero-bias conductance peak at the edge
of an InSb quantum wire (QW) coupled to an $s$-wave superconducting
substrate, which might be the first experimental observation of a
Majorana fermion~\cite{Science.336.1003}. It has also been shown
theoretically that a one-dimensional Rashba quantum wire coupled to a
$d$-wave superconductor hosts doubly degenerate Majorana bound states
(MBSs) at the edge~\cite{arXiv:1211.0338}.
Besides these distinct works, there are various other research
activities on the basis of QWs (for example
Refs.~\cite{PhysRevB.82.214509, PhysRevLett.105.077001,
PhysRevB.83.094525, PhysRevLett.105.177002, NewJPhys.13.065004,
PhysRevB.84.144522, PhysRevLett.106.127001, arXiv:1208.3701}).
In all the cases listed above, spin-orbit interaction and/or the Zeeman
energy in the QW is essential. Therefore, an open question is if zero
energy Majorana bound states are also realizable in a one-dimensional
system in the absence of these interactions, which opens new
opportunities as we will show below.

In this paper, we investigate electronic states caused by the proximity
effect between a metallic QW and a two-dimensional (2D) unconventional
superconductor to search for zero energy MBSs at the ends of the QW.
We discover that the resulting MBSs are doubly degenerate, i.e.,
characterized by a spin degree of freedom. This leads to nonlocal
spin-correlations between the two ends of the QW which can be
manipulated by all-electric means.
For a substrate in symmetry class D~\cite{PhysRevB.78.195125,
*NewJPhys.12.065010}, i.e., in the absence of time reversal symmetry
(TRS) the degenerate end states can be gapped out by switching on TRS
preserving local imperfections that couple opposite spin. This is
reflected in the fact these systems are characterized by a trivial
$\mathbb Z_2$~topological invariant. In contrast, for a TRS preserving
substrate in symmetry class DIII, we find helical MBS pairs that are
topologically protected by TRS.

\begin{figure}
 \begin{center}
  \includegraphics[width=.5\hsize]{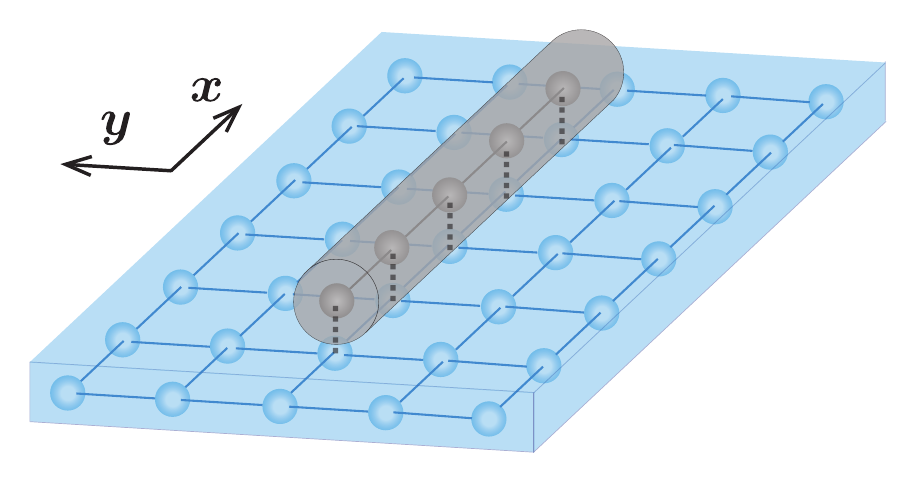}%
  \caption{\label{fig:system}One dimensional metallic quantum wire (QW)
  on top of a substrate of an unconventional two dimensional SC. The
  wire is located along the $x$ axis. Dashed lines illustrate tunneling
  between the QW and the SC, and solid spheres represent the sites of
  our tight-binding model discussed in the text.}
 \end{center}
\end{figure}

\textit{Model.--}
Our setup is shown schematically in Fig.~\ref{fig:system} and the model
Hamiltonian given by
\begin{align}
H_{\mathrm{SC}}& =
 \int\mathrm{d}^2 k \Psi_{\bm{k}}^{\dagger}
 \begin{pmatrix}
  \frac{k_x^2+k_y^2}{2m_s}-\mu_s & \hat{\Delta} \\
  \hat{\Delta}^{\dagger} & -\frac{k_x^2+k_y^2}{2m_s}+\mu_s \\
 \end{pmatrix}
 \Psi_{\bm{k}},
\label{eq:pSC}
\end{align}
\begin{align}
\Psi_{\bm{k}}& =\left( a_{\uparrow\bm{k}}\; a_{\downarrow\bm{k}}\;
 a_{\uparrow\bm{-k}}^{\dagger}\; a_{\downarrow\bm{-k}}^{\dagger}
 \right),
\\
H_{\mathrm{wire}}&=
 \int\mathrm{d}k\psi_{\sigma k}^{\dagger}\left[
 \frac{k_x^2}{2m}-\mu
 \right]\psi_{\sigma k},
\label{eq:wire}
\end{align}
\begin{align}
H_{\mathrm{\mathrm{t}}}&=
 c \int \mathrm{d}x
 \left( \psi^{\dagger}_{x \sigma} a_{(x,0) \sigma} + \mathrm{h.c.}\right),
\label{eq:tunnel}
\end{align}
where $H_{\mathrm{wire}}$ describes the QW put along $x$-direction on the
surface of the superconductor described by $H_{\mathrm{SC}}$ which
infinitely spreads in the $xy$-plane.
These two are connected by a tunneling Hamiltonian $H_{\mathrm{t}}$ with the
spin-independent constant matrix element $c$.
We consider the following three cases for the order parameter matrix
$\hat{\Delta}$:
(i) $\hat{\Delta}=i\sigma_{y}\Delta\sigma_{z}( k_x + i k_y)$ for the chiral
superconductor of class D with the topological invariant $Z=1$, and
(ii) $\hat{\Delta}=i\sigma_{y}\Delta\bm{k}\cdot\bm{\sigma}$ for the time
reversal symmetric class DIII superconductor with the topological
invariant $\mathbb Z_2=1$. Here, the $\sigma$~denotes Pauli matrices in
spin space. As the third case (iii), we consider a system where two
layers of a 2D superconductor of case (ii) are coupled to each other by
an interlayer transfer integral, which is a topologically trivial case
with $\mathbb Z_2=0$.
The model (i) is relevant to Sr$_2$RuO$_4$ where the chiral $p+ip$
superconductivity with broken time-reversal symmetry
is believed to be realized~\cite{RevModPhys.75.657}.
The model (ii) is the 2D analogue of the He B phase, and is continuously
connected to the helical non-centrosymmetric topological superconductor with
Rashba spin-orbit interaction~\cite{PhysRevB.79.060505,
PhysRevB.79.094504, PhysRevLett.108.147003}.
We present the calculational steps and results mainly for case (i)
below, but it is straightforward to apply them to the other two cases
(ii) and (iii).
Our main purpose is to investigate the electronic states in the
one-dimensional superconductivity formed in this way and to discuss its
topological nature as well as resulting Majorana bound states.

\begin{figure}
 \begin{center}
  \includegraphics[width=.8\hsize]{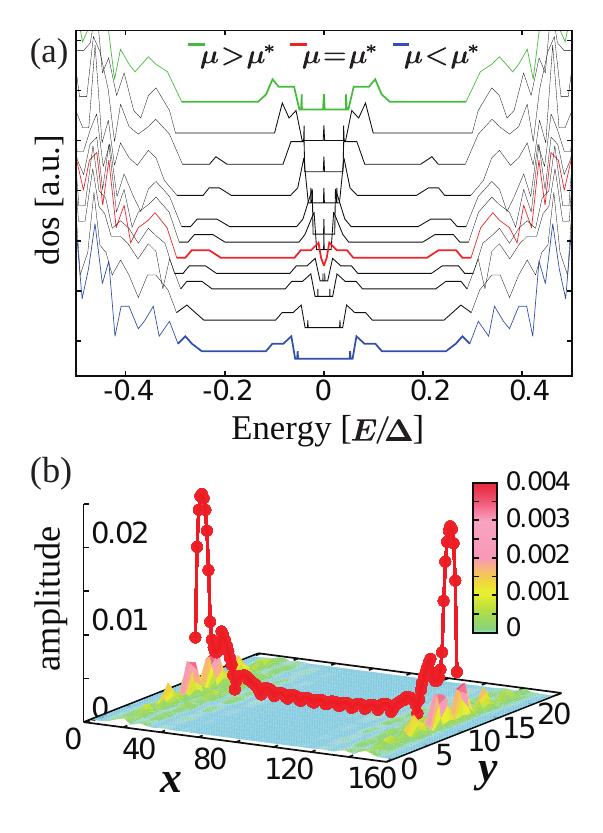}%
  \caption{\label{fig:spectrum}(a) Density of states with different
  values of $\mu$. The horizontal axis is in units of the substrate
  superconducting gap $\Delta$. This plot shows the topological phase
  transition at $\mu=\mu^{*}$. There are no states inside the gap when
  $\mu<\mu^{*}$ while there are four-fold degenerate zero energy states
  when $\mu>\mu^{*}$. (b) Probability distribution of the zero energy
  states. The size of system is $160\times24$ for the substrate and
  $140\times1$ for the wire. Red solid circles show the probability
  distribution in the wire, which has peaks at both ends.}
 \end{center}
\end{figure}

\textit{Effective Green's function.--}
First, we derive the effective Green's function of the QW by
integrating over the superconductor
\begin{align}
 G_{\mathrm{eff}}^{-1}(i\omega_{n},k_x)&=
 i\omega_{n}-H_{\mathrm{wire}}(k_x)\tau_z - c^2 I(i\omega,k_x),
 \label{eq:geff}
 \\
 I(i\omega_{n},k_x)&=
 \int \mathrm{d}k_y
 \frac{i\omega_{n} + H_{\mathrm{SC}}(k_x,k_y)}
 {(i\omega_{n})^2 - H_{\mathrm{SC}}^2(k_x,k_y)},
 \label{eq:I}
\end{align}
where $\tau$~denotes Pauli matrices in the Nambu space.
As long as there is neither spin-orbit interaction in the wire nor
spin-dependent tunneling between the wire and the superconductor,
theinduced superconducting order parameter is also spin-triplet and can
be decoupled into two sectors $\sigma_x=\pm1$.
The integration can be analytically performed and we obtain
\begin{align}
 & G_{\mathrm{eff}}^{-1}(i\omega_{n},k_x)=
 i\omega_{n}(1-2c^2I_1)-\bm{h}(k_x) \cdot \bm{\tau},
 \\
 & \bm{h}(k_x)=
 \left( 2c^2\Delta k_x I_1\sigma_x,\; 0,\;
 \xi_k+2c^2\xi_k^{\mathrm{SC}}I_1+2c^2I_2 \right),
 \label{eq:h}
 \\
 & I_1=
 \frac{ m_{\mathrm{s}} \left[ \lambda_+^{-1} \arctan \left(
 \kappa/\lambda_+ \right) - \lambda_-^{-1} \arctan \left( \kappa/\lambda_-
 \right) \right]}
 {2\pi \sqrt{m_{\mathrm{s}}^2\Delta^4 -
 2m_{\mathrm{s}}\Delta^2\mu_{\mathrm{s}} - \omega_{n}^2}},
 \\
 & I_2=
 \frac{\lambda_- \arctan \left( \kappa/\lambda_-\right) - \lambda_+
 \arctan \left( \kappa/\lambda_+ \right)}
 {4\pi\sqrt{m_{\mathrm{s}}^2\Delta^4 -
 2m_{\mathrm{s}}\Delta^2\mu_{\mathrm{s}} - \omega_{n}^2}},
\end{align}
where $\kappa$ is a momentum cut-off and
$\lambda_{\pm} = \left( k_x^2 + 2m_{\mathrm{s}}^2\Delta^2
- 2m_{\mathrm{s}}\mu_{\mathrm{s}} \pm 2m_{\mathrm{s}}\sqrt{m_{\mathrm{s}}^2\Delta^4 -
2m_{\mathrm{s}}\Delta^2\mu_{\mathrm{s}} - \omega_{n}^2} \right)^{1/2}$,
$\xi_k=k^2/2m-\mu$, $\xi_k^{\mathrm{SC}}$ likewise.
In the following, we look at the $\sigma_x=1$ sector, which is
equivalent to a spinless $p$-wave superconductor.
From Eq.~(\ref{eq:h}), we can then calculate the spectral function
$A(\omega, k_x)$ of the electron as
\begin{align}
 A(\omega,k_x)=-{ 1 \over {2\pi} } \Im\left[ \mathrm{Tr} \left[ \tau_z
 G_{\mathrm{eff}}^{\mathrm{R}}(\omega,k_x)\right] \right],
 \label{eq:sf}
\end{align}
where $G_{\mathrm{eff}}^{\mathrm{R}}(\omega,k_x)$ is the retarded
Green's function obtained from $G_{\mathrm{eff}}(i\omega_{n},k_x)$~via the
analytic continuation $i\omega_{n} \to \omega+i\delta$, where $\delta$~is a
infinitesimally small positive number.
\begin{figure}
 \begin{center}
  \includegraphics[width=.9\hsize]{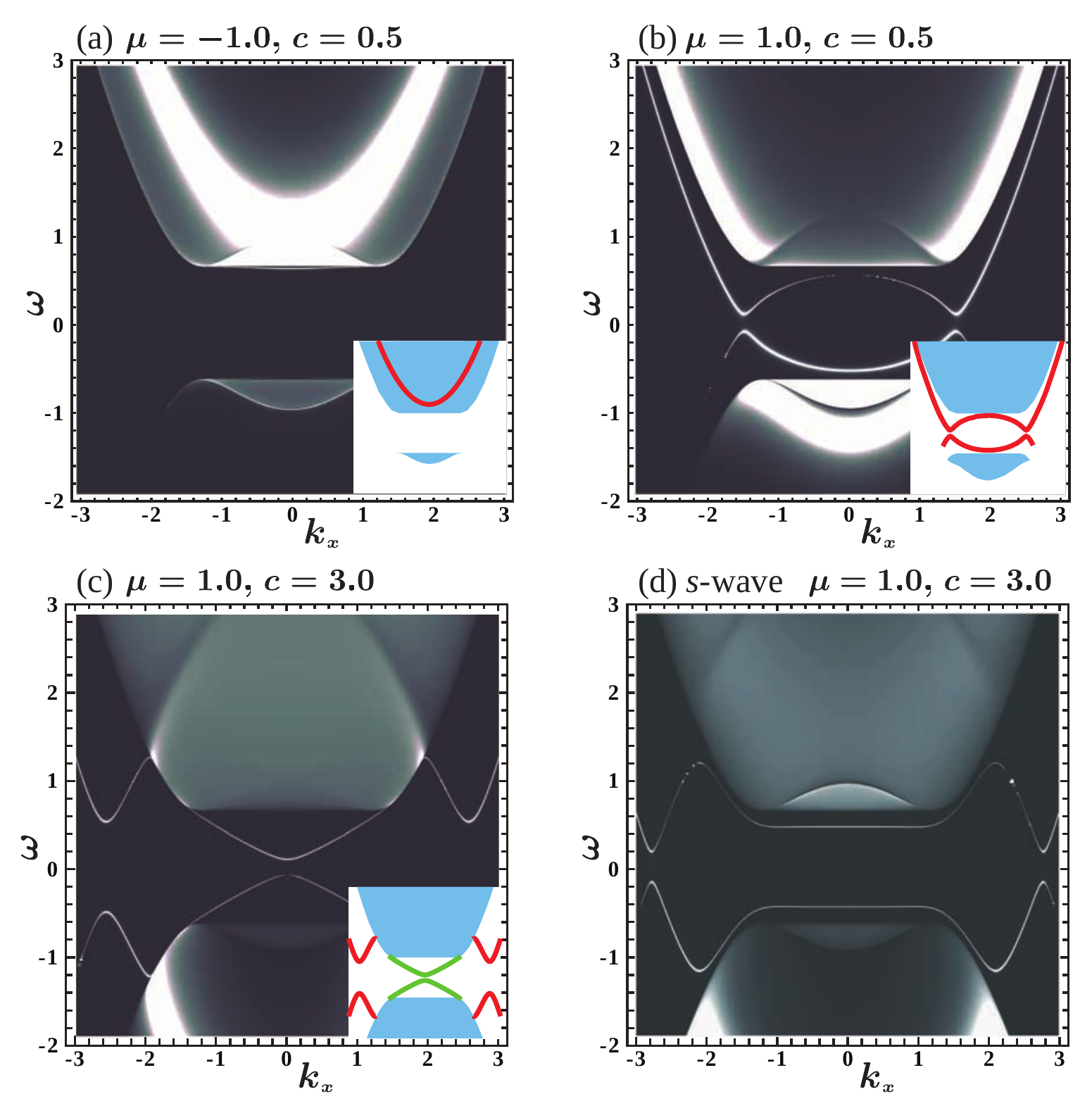}%
  \caption{\label{fig:sf}Spectral function of the electronic
  excitations. The insets show the schematic structure of the spectrum;
  Blue: continuum of the substrate superconductor, Red: 1D states of the
  QW, and Green: edge states of the 2D SC induced by the coupling to the
  QW.
  In (a) the states in the QW are located outside the superconducting
  gap, while in (b) these states are within the superconducting gap and
  an SC order parameter is induced in the QW. (c) is the case where the
  coupling is strong enough that the bonding and anti-bonding states act
  as a potential barrier and edge states in the 2D superconductor
  emerge. (d) shows that no zero energy states are formed in an $s$-wave
  superconductor under the same conditions as (c).}
 \end{center}
\end{figure}
In Fig.~\ref{fig:sf}, we plot Eq.~(\ref{eq:sf}) with $m=m_s=1.0,\,
\mu_s=1.0,\, \Delta=0.5$, and several values of $\mu$~and $c$.
The insets show the schematic structure of the spectrum. The red lines
represent states in the wire without and with an induced superconducting
gap, in cases (a) and (b) respectively. The blue shaded area shows the
continuum of the bulk states in the superconductor, which arises in the
effective Green's function of the QW as the self-energy
(Eqs.~(\ref{eq:geff})~and~(\ref{eq:I})). The green lines are edge states
in the 2D SC that will be further discussed below.
Figures~\ref{fig:sf}(a) and (b) correspond to the cases where the band
of the QW is outside and inside of the superconducting gap of the
substrate superconductor, respectively. We can see that in
Fig.~\ref{fig:sf}(b) a finite gap is induced by the proximity effect. An
interesting situation arises when the magnitude of the transfer integral
related to the tunneling $c$~is strong enough that bonding and
anti-bonding states are formed at the interface of the wire and the
superconductor. The energy of those states is pushed away from the low
energy regime, and they act as a potential barrier to other
states. Therefore, a boundary is effectively induced in the substrate
superconductor. It is known that edge modes appear at the boundary of 2D
topological superconductors. The energy of the midgap states around
$k_x=0$ approach zero as the magnitude of $c$ becomes larger (see
Fig.~\ref{fig:sf}(c)). This shows the formation of edge channel that are
gapped out because the edge modes located at both sides of the wire
weakly interact with each other. In Fig.~\ref{fig:sf}(d) we show the
spectral function for an $s$-wave superconductor substrate for
comparison. There are no states corresponding to the edge channel in
case (c).
\textit{Topological nature.--}
Now, we focus on the topological nature of the QW system. For this
purpose, we use the method for the calculation of the topological number
in terms of Green's function at zero frequency~\cite{PhysRevX.2.031008,
arXiv:1207.1104}. We derive an effective Hamiltonian from
Eq.~(\ref{eq:geff}) on the basis of
$H_{\mathrm{eff}}(k_x)=-G_{\mathrm{eff}}^{-1}(i\omega_{n}=0,k_x) =
\bm{h}(k_x)\cdot\bm{\tau}$ for a given $\sigma_x = \pm 1$. This
2$\times$2 Hamiltonian is similar to that of Kitaev's original proposal
of a 1D topological superconductor~\cite{PhysUsp.44.131}. It is known
that the system is in the topological phase when
$\hat{\bm{h}}(k_x)=\bm{h}/|\bm{h}|$ connects antipodal points of the
unit sphere as $k_x$ is varied from the center to the boundary of the
Brillouin zone \cite{RepProgPhys.75.076501}. Therefore, it is easy to
verify that when $\mu > 2c^2 (-\mu_{\mathrm{s}} I_1(k_x=0) +
I_2(k_x=0))=\mu^* $, the QW becomes two copies of Kitaev's 1D
topological superconductor with $\mathbb Z_2=1$.
The topological nature of the 1D superconductor manifests
itself in bound states at the ends of the QW, which will benumerically
studied in the following. However, in case (i), the two identical copies
of the $\mathbb Z_2$-nontrivial Kitaev model form a trivial composite
system. Physically, this implies that the degenerate end states can be
gapped out by switching on terms that couple the opposite spin sectors
as we confirmed by numerical calculations.
\textit{Numerical study of a tight-binding model and Majorana bound
states.-} We study a tight-binding model by numerical calculations. We
change our continuum model to a lattice model, i.e., $k \to \sin k$ and
$k^2 \to 2(1-\cos k)$ in Eq.~(\ref{eq:h}), and construct the
corresponding tight-binding Hamiltonian. Then, we calculate the energy
spectrum in the system with \textit{open boundary conditions} as
depicted in Fig.~\ref{fig:system}. Figure~\ref{fig:spectrum}(a) shows
the topological phase transition as a function of the chemical potential
in the QW.
In the strong coupling regime, where $\mu<\mu^{*}$, there are no states
in the induced superconducting gap, while in the weak coupling regime,
where $\mu>\mu^{*}$ there are zero energy states. It can be checked that
four-fold degenerate zero energy states are formed, two at each
end. Fig.~\ref{fig:spectrum}(b) shows the probability distributions of
one of them. Red solid circles represent the weight in the QW, and we
can see the state is localized at the ends of the wire. In sharp
contrast to the Kitaev model, we have two spin sectors in the QW which
are degenerate. This fact attaches a spin degree of freedom to the
MBSs. However, in the presence of spin-orbit interaction, which mixes
the $\sigma_x=\pm 1$ sectors, the degeneracy is lifted and the Majorana
fermions will be pushed away from zero energy.

\begin{figure}
 \begin{center}
  \includegraphics[width=.7\hsize]{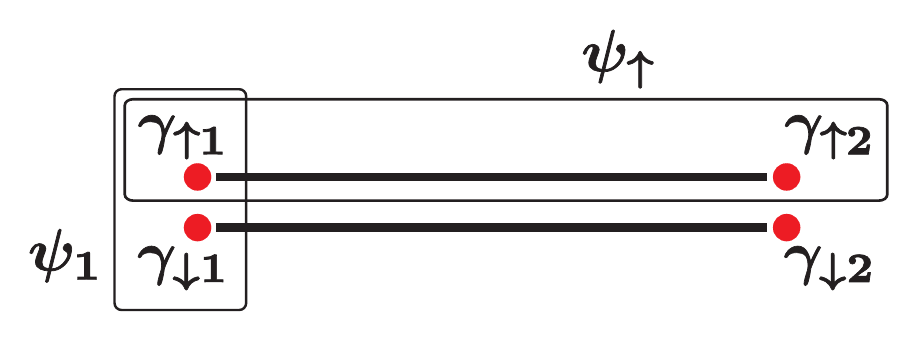}%
  \caption{\label{fig:majorana}Schematic figure of Majorana states at the
  ends of the wire (red circles). As discussed in the main text, the QW
  can be regarded as two copies of Kitaev model (solid lines). There are
  two ways to combine two of them into ordinary fermions:
  \textit{non-local} and \textit{local} ones.}
 \end{center}
\end{figure}

In the following, we briefly discuss the other two possibilities (ii)
and (iii) for the SC substrate. We have confirmed that a similar
analysis to case (i) outlined above applies to these cases, and we have
found that the doubly degenerate zero energy Majorana bound states
appear when the chemical potential $\mu$ of the QW is positive and large
enough for the topological phase transition to occur. In cases (ii) and
(iii), the spin degenerate pair of MBS stems from a non-trivial $\mathbb
Z_2$~invariant for the composite system in symmetry class
DIII. Therefore, these end states are topologically protected by TRS and
cannot be gapped out by switching on spin-orbit interaction as long as
the bulk gap is maintained. This is an essential difference to case (i)
where the conservation of $\sigma_x=\pm 1$ was needed in addition to the
generic symmetries of the model to obtain zero energy MBSs, while TRS
protects the doubly degenerate MBSs in cases (ii) and (iii). This is
particularly remarkable for case (iii) where the substrate 2D
superconductor is originally in a topologically trivial phase.
\textit{Local and non-local fermions.--}
There have been many theoretical proposals to use chiral MBSs for
topological quantum computing~\cite{PhysRevB.84.094505,
PhysRevB.84.144509,PhysRevB.85.144501}. Chiral MBSs are difficult to
detect and to manipulate because they only couple in a limited way to
their environment. This is different in our proposal. On the one hand,
the doubly degenerate MBSs discussed above are more susceptible
regarding their coupling to the environment. On the other hand, they can
show novel responses through a strong coupling between the fermion
parity and the spin degree of freedom. The idea behind is schematically
shown in Fig.~\ref{fig:majorana}. The solid lines represent Kitaev
chains with MBSs indicated by red circles. Let us elaborate a bit more
on the two kinds of fermions that can be formed by our MBSs.
Since Majorana fermions are real, normal fermions are formed from pairs
of them. We can construct two kinds of normal (complex) fermions,
\textit{non-local} spin polarized ones
$\psi_{\uparrow}=1/2(\gamma_{\uparrow 1}+i\gamma_{\uparrow 2})$ and
$\psi_{\downarrow}=1/2(\gamma_{\downarrow 1}-i\gamma_{\downarrow 2})$,
and \textit{local} ones
$\psi_{1}=1/2(\gamma_{\uparrow 1}+i\gamma_{\downarrow 1})$ and
$\psi_{2}=1/2(\gamma_{\uparrow 2}-i\gamma_{\downarrow 2})$.
The local pseudo spin operators are defined as
$\bm{s}_{1(2)}=1/2 \times \gamma_{1(2)\alpha}\bm{\sigma}_{\alpha\beta}\gamma_{1(2)\beta}$.
It is known that Majorana fermions have only Ising-type spin due to
their anticommutation relation
$\{\gamma_{i},\gamma_{j}\}=2\delta_{ij}$, then, $s_x=s_z=0$ and
$s_y=1/2-\psi^{\dagger}\psi$ from the above definition.
We would like to emphasize that these local variables are specific to
the paired Majorana states. There will be various interesting physics as
a consequence of the synthesis of the local and the non-local
nature. Here, as an example, we focus on spin-spin correlation of two
Ising spins formed at the ends.
In preparation for that, we first review some aspects of the fermion parity.
When we have two Majorana operators, $\gamma$ and $\gamma^{\prime}$,
there exist two orthogonal states $|0\rangle$ and $|1\rangle$, which
satisfy $-i\gamma\gamma^{\prime}|0\rangle=|0\rangle$~and
$-i\gamma\gamma^{\prime}|1\rangle=-|1\rangle$, respectively. The former
has an even fermion parity and the latter odd.
They can be rewritten by normal fermion operators defined as
$\psi=1/2(\gamma+i\gamma^{\prime})$ where $\psi^{\dagger}\psi|0\rangle=0$ and
$\psi^{\dagger}\psi|1\rangle=|1\rangle$.
Note that if the wire is connected to ground by a capacitor, one can
modify the fermion number via charging energy $U(n)=Q^2/(2C)- V(t)Q=
(Q-Q_0)^2/(2C) + \mathrm{const.}$ where $Q=ne$ is the charge, $V(t)$ is
the gate voltage, and $Q_0 = CV(t)$.
The importance of charging energy in the context of topological
superconductors was first discussed in
Ref.~\cite{PhysRevLett.104.056402}, where electron
teleportation (mediated by the non-locality of the MBSs) has been proposed.
We now apply this idea to our model.
Our Hilbert space is 4-dimensional and there are
two kinds of bases one can construct, i.e., from $\psi_1$, $\psi_2$, or from
$\psi_{\uparrow}$, and $\psi_{\downarrow}$.
Namely, one can define
$\psi_1^{\dagger}\psi_1^{\phantom{\dagger}}|1\rangle_{\!1}=|1\rangle_{\!1}$,
\textit{etc.}, and
$|0\rangle_{\!1}$, $|1\rangle_{\!1}$, \dots are local basis states while
$|0\rangle_{\uparrow}$, $|1\rangle_{\uparrow}$,  \dots are non-local
basis states. The relation between these two kinds of basis states is given by
\begin{align}
\begin{pmatrix}
|0\rangle_{\uparrow} |0\rangle_{\downarrow} \\
|0\rangle_{\uparrow} |1\rangle_{\downarrow} \\
|1\rangle_{\uparrow} |0\rangle_{\downarrow} \\
|1\rangle_{\uparrow} |1\rangle_{\downarrow}
\end{pmatrix}
=
\frac{1}{\sqrt{2}}
\begin{pmatrix}
 0 & -i & 1 &  0 \\
 1 &  0 & 0 &  i \\
 1 &  0 & 0 & -i \\
 0 &  i & 1 &  0
\end{pmatrix}
\begin{pmatrix}
|0\rangle_{1} |0\rangle_{2} \\
|0\rangle_{1} |1\rangle_{2} \\
|1\rangle_{1} |0\rangle_{2} \\
|1\rangle_{1} |1\rangle_{2}
\end{pmatrix}.
\end{align}
In our system, the total number of fermions can be controlled by the
gate voltage $V(t)$.
We consider $s_1 s_2$ correlation in the odd and even fermion number
space, respectively. The density matrices are
$\rho_{\mathrm{odd}}=
\frac{1}{2}\left(|1\rangle_{\!1} |0\rangle_{\!2} {}_{1\!}\langle0| {}_{2\!}\langle1|
+|0\rangle_{\!1} |1\rangle_{\!2} {}_{1\!}\langle1| {}_{2\!}\langle0|\right)$
and
$\rho_{\mathrm{even}}=
\frac{1}{2}\left(|0\rangle_{\!1} |0\rangle_{\!2} {}_{1\!}\langle0| {}_{2\!}\langle0|
+|1\rangle_{\!1} |1\rangle_{\!2} {}_{1\!}\langle1| {}_{2\!}\langle1|\right)$.
Then we obtain
\begin{align}
- \langle s_{1y}s_{2y} \rangle_{\mathrm{odd}}
= \langle s_{1y}s_{2y} \rangle_{\mathrm{even}}
=\frac{1}{4}\,\,,
\end{align}
where $\langle\dots\rangle_{\mathrm{odd}/\mathrm{even}}$ represents the
statistical average with the density matrices $\rho_{\mathrm{odd}}$~and $\rho_{\mathrm{even}}$, respectively.
These results show that by modulating charging energy we can control
spin-spin correlation, which is independent of the distance between the
two spins because they are connected through the superconducting wire in
between.
This effect is a consequence of both, the local and the non-local
degrees of freedom of the system.
One can also manipulate the two Ising spins by modulating the
gate voltage as a function of time as $V(t) = V_0 + \delta V(t)$.
The equations of motion of $s_{1y}$ and $s_{2y}$ by the driving force
$\delta V(t) (i\gamma_{1\uparrow}\gamma_{2\uparrow} +
i\gamma_{1\downarrow}\gamma_{2\downarrow})$ yield
\begin{align}
\frac{\mathrm{d}}{\mathrm{d} t}(s_{1y}+s_{2y})&=0
\label{eq:eom1}
\\
\frac{\mathrm{d}^2}{\mathrm{d}^2 t}(s_{1y}-s_{2y}) &= -4\left(\frac{2e
 \delta V_0}{\hbar}\right)^2(s_{1y}-s_{2y})
\label{eq:eom2}
\end{align}
for a constant $\delta V_0$ during some time period $T$.
By tuning $T$ to satisfy $4e \delta V_0T/\hbar = \pi$ one can reverse
the pseudospins $s_y$ components when they are anti-parallel while they
remain unchanged when they are parallel.
\vspace{10pt}
\textit{Conclusion.--}
In this paper, we have studied theoretically the proximity effect of a
one-dimensional metallic quantum wire without spin-orbit interaction on
the substrate of an unconventional superconductor. We have considered
three different cases for the substrate: (i) a chiral superconductor;
(ii) a class DIII superconductor with a non-trivial $\mathbb Z_2$
number, and (iii) a class DIII superconductor with a trivial $\mathbb
Z_2$ number. Interestingly, we have found degenerate zero energy
Majorana bound states at both ends of the quantum wire for all three
cases. We have shown that these degenerate Majorana bound states for
cases (ii) and (iii) can be nicely used for a combined spin/parity
qubit. Furthermore, we have demonstrated that the resulting non-local
spin correlations between the two ends of the wire can be controlled by
the gate voltage potential acting on the wire. One of the candidate
systems for the case (ii) substrate are thin films of
Sr$_{2}$RuO$_{4}$~\cite{NewJPhys.11.055070} or bi-layer Rashba
system~\cite{PhysRevLett.108.147003}, where the class DIII
superconducting state might be realized, and for the case (iii) super
lattice structure of CeCoIn$_5$~\cite{NatPhys.7.849}.

\begin{acknowledgments}
\textit{Acknowledgment.-}SN was supported by Grant-in-Aid for JSPS
 Fellows and JCB by the Swedish Research Council. This work was further
 supported by Grant-in-Aid for Scientific Research (S) (Grants
 No. 24224009) from the Ministry of Education, Culture, Sports, Science
 and Technology of Japan, Strategic International Cooperative Program
 (Joint Research Type) from Japan Science and Technology Agency, Funding
 Program for World-Leading Innovative RD on Science and Technology
 (FIRST Program), the Deutsche Forschungsgemeinschaft, the European
 Science Foundation, and the Helmholtz Foundation.
\end{acknowledgments}

%


\begin{thebibliography}{40}%
\makeatletter
\providecommand \@ifxundefined [1]{%
 \@ifx{#1\undefined}
}%
\providecommand \@ifnum [1]{%
 \ifnum #1\expandafter \@firstoftwo
 \else \expandafter \@secondoftwo
 \fi
}%
\providecommand \@ifx [1]{%
 \ifx #1\expandafter \@firstoftwo
 \else \expandafter \@secondoftwo
 \fi
}%
\providecommand \natexlab [1]{#1}%
\providecommand \enquote  [1]{``#1''}%
\providecommand \bibnamefont  [1]{#1}%
\providecommand \bibfnamefont [1]{#1}%
\providecommand \citenamefont [1]{#1}%
\providecommand \href@noop [0]{\@secondoftwo}%
\providecommand \href [0]{\begingroup \@sanitize@url \@href}%
\providecommand \@href[1]{\@@startlink{#1}\@@href}%
\providecommand \@@href[1]{\endgroup#1\@@endlink}%
\providecommand \@sanitize@url [0]{\catcode `\\12\catcode `\$12\catcode
  `\&12\catcode `\#12\catcode `\^12\catcode `\_12\catcode `\%12\relax}%
\providecommand \@@startlink[1]{}%
\providecommand \@@endlink[0]{}%
\providecommand \url  [0]{\begingroup\@sanitize@url \@url }%
\providecommand \@url [1]{\endgroup\@href {#1}{\urlprefix }}%
\providecommand \urlprefix  [0]{URL }%
\providecommand \Eprint [0]{\href }%
\providecommand \doibase [0]{http://dx.doi.org/}%
\providecommand \selectlanguage [0]{\@gobble}%
\providecommand \bibinfo  [0]{\@secondoftwo}%
\providecommand \bibfield  [0]{\@secondoftwo}%
\providecommand \translation [1]{[#1]}%
\providecommand \BibitemOpen [0]{}%
\providecommand \bibitemStop [0]{}%
\providecommand \bibitemNoStop [0]{.\EOS\space}%
\providecommand \EOS [0]{\spacefactor3000\relax}%
\providecommand \BibitemShut  [1]{\csname bibitem#1\endcsname}%
\let\auto@bib@innerbib\@empty
\bibitem [{\citenamefont {Fu}\ and\ \citenamefont
  {Kane}(2008)}]{PhysRevLett.100.096407}%
  \BibitemOpen
  \bibfield  {author} {\bibinfo {author} {\bibfnamefont {L.}~\bibnamefont
  {Fu}}\ and\ \bibinfo {author} {\bibfnamefont {C.~L.}\ \bibnamefont {Kane}},\
  }\href {\doibase 10.1103/PhysRevLett.100.096407} {\bibfield  {journal}
  {\bibinfo  {journal} {Phys. Rev. Lett.}\ }\textbf {\bibinfo {volume} {100}},\
  \bibinfo {pages} {096407} (\bibinfo {year} {2008})}\BibitemShut {NoStop}%
\bibitem [{\citenamefont {Beenakker}(2011)}]{arXiv:1112.1950}%
  \BibitemOpen
  \bibfield  {author} {\bibinfo {author} {\bibfnamefont {C.~W.~J.}\
  \bibnamefont {Beenakker}},\ }\href {http://arxiv.org/abs/1112.1950v2}
  {\bibfield  {journal} {\bibinfo  {journal} {arXiv:1112.1950}\ } (\bibinfo
  {year} {2011})}\BibitemShut {NoStop}%
\bibitem [{\citenamefont {Tanaka}\ \emph {et~al.}(2012)\citenamefont {Tanaka},
  \citenamefont {Sato},\ and\ \citenamefont {Nagaosa}}]{JPSJ.81.011013}%
  \BibitemOpen
  \bibfield  {author} {\bibinfo {author} {\bibfnamefont {Y.}~\bibnamefont
  {Tanaka}}, \bibinfo {author} {\bibfnamefont {M.}~\bibnamefont {Sato}}, \ and\
  \bibinfo {author} {\bibfnamefont {N.}~\bibnamefont {Nagaosa}},\ }\href
  {\doibase 10.1143/JPSJ.81.011013} {\bibfield  {journal} {\bibinfo  {journal}
  {J. Phys. Soc. Jpn.}\ }\textbf {\bibinfo {volume} {81}},\ \bibinfo {pages}
  {011013} (\bibinfo {year} {2012})}\BibitemShut {NoStop}%
\bibitem [{\citenamefont {Alicea}(2012)}]{RepProgPhys.75.076501}%
  \BibitemOpen
  \bibfield  {author} {\bibinfo {author} {\bibfnamefont {J.}~\bibnamefont
  {Alicea}},\ }\href {http://stacks.iop.org/0034-4885/75/i=7/a=076501}
  {\bibfield  {journal} {\bibinfo  {journal} {Rep. Prog. Phys.}\ }\textbf
  {\bibinfo {volume} {75}},\ \bibinfo {pages} {076501} (\bibinfo {year}
  {2012})}\BibitemShut {NoStop}%
\bibitem [{\citenamefont {Majorana}(1937)}]{NuovoCim.14.171}%
  \BibitemOpen
  \bibfield  {author} {\bibinfo {author} {\bibfnamefont {E.}~\bibnamefont
  {Majorana}},\ }\href@noop {} {\bibfield  {journal} {\bibinfo  {journal}
  {Nuovo Cimento}\ }\textbf {\bibinfo {volume} {14}},\ \bibinfo {pages} {171}
  (\bibinfo {year} {1937})}\BibitemShut {NoStop}%
\bibitem [{\citenamefont {Wilczek}(2009)}]{NatPhys.5.614}%
  \BibitemOpen
  \bibfield  {author} {\bibinfo {author} {\bibfnamefont {F.}~\bibnamefont
  {Wilczek}},\ }\href {\doibase 10.1038/nphys1380} {\bibfield  {journal}
  {\bibinfo  {journal} {Nature Phys.}\ }\textbf {\bibinfo {volume} {5}},\
  \bibinfo {pages} {614} (\bibinfo {year} {2009})}\BibitemShut {NoStop}%
\bibitem [{\citenamefont {Franz}(2010)}]{Physics.3.24}%
  \BibitemOpen
  \bibfield  {author} {\bibinfo {author} {\bibfnamefont {M.}~\bibnamefont
  {Franz}},\ }\href {\doibase 10.1103/Physics.3.24} {\bibfield  {journal}
  {\bibinfo  {journal} {Physics}\ }\textbf {\bibinfo {volume} {3}},\ \bibinfo
  {pages} {24} (\bibinfo {year} {2010})}\BibitemShut {NoStop}%
\bibitem [{\citenamefont {Read}\ and\ \citenamefont
  {Green}(2000)}]{PhysRevB.61.10267}%
  \BibitemOpen
  \bibfield  {author} {\bibinfo {author} {\bibfnamefont {N.}~\bibnamefont
  {Read}}\ and\ \bibinfo {author} {\bibfnamefont {D.}~\bibnamefont {Green}},\
  }\href {\doibase 10.1103/PhysRevB.61.10267} {\bibfield  {journal} {\bibinfo
  {journal} {Phys. Rev. B}\ }\textbf {\bibinfo {volume} {61}},\ \bibinfo
  {pages} {10267} (\bibinfo {year} {2000})}\BibitemShut {NoStop}%
\bibitem [{\citenamefont {Ivanov}(2001)}]{PhysRevLett.86.268}%
  \BibitemOpen
  \bibfield  {author} {\bibinfo {author} {\bibfnamefont {D.~A.}\ \bibnamefont
  {Ivanov}},\ }\href {\doibase 10.1103/PhysRevLett.86.268} {\bibfield
  {journal} {\bibinfo  {journal} {Phys. Rev. Lett.}\ }\textbf {\bibinfo
  {volume} {86}},\ \bibinfo {pages} {268} (\bibinfo {year} {2001})}\BibitemShut
  {NoStop}%
\bibitem [{\citenamefont {Das~Sarma}\ \emph {et~al.}(2005)\citenamefont
  {Das~Sarma}, \citenamefont {Freedman},\ and\ \citenamefont
  {Nayak}}]{PhysRevLett.94.166802}%
  \BibitemOpen
  \bibfield  {author} {\bibinfo {author} {\bibfnamefont {S.}~\bibnamefont
  {Das~Sarma}}, \bibinfo {author} {\bibfnamefont {M.}~\bibnamefont {Freedman}},
  \ and\ \bibinfo {author} {\bibfnamefont {C.}~\bibnamefont {Nayak}},\ }\href
  {\doibase 10.1103/PhysRevLett.94.166802} {\bibfield  {journal} {\bibinfo
  {journal} {Phys. Rev. Lett.}\ }\textbf {\bibinfo {volume} {94}},\ \bibinfo
  {pages} {166802} (\bibinfo {year} {2005})}\BibitemShut {NoStop}%
\bibitem [{\citenamefont {Nayak}\ \emph {et~al.}(2008)\citenamefont {Nayak},
  \citenamefont {Simon}, \citenamefont {Stern}, \citenamefont {Freedman},\ and\
  \citenamefont {Das~Sarma}}]{RevModPhys.80.1083}%
  \BibitemOpen
  \bibfield  {author} {\bibinfo {author} {\bibfnamefont {C.}~\bibnamefont
  {Nayak}}, \bibinfo {author} {\bibfnamefont {S.~H.}\ \bibnamefont {Simon}},
  \bibinfo {author} {\bibfnamefont {A.}~\bibnamefont {Stern}}, \bibinfo
  {author} {\bibfnamefont {M.}~\bibnamefont {Freedman}}, \ and\ \bibinfo
  {author} {\bibfnamefont {S.}~\bibnamefont {Das~Sarma}},\ }\href {\doibase
  10.1103/RevModPhys.80.1083} {\bibfield  {journal} {\bibinfo  {journal} {Rev.
  Mod. Phys.}\ }\textbf {\bibinfo {volume} {80}},\ \bibinfo {pages} {1083}
  (\bibinfo {year} {2008})}\BibitemShut {NoStop}%
\bibitem [{\citenamefont {Linder}\ \emph {et~al.}(2010)\citenamefont {Linder},
  \citenamefont {Tanaka}, \citenamefont {Yokoyama}, \citenamefont {Sudb\o{}},\
  and\ \citenamefont {Nagaosa}}]{PhysRevLett.104.067001}%
  \BibitemOpen
  \bibfield  {author} {\bibinfo {author} {\bibfnamefont {J.}~\bibnamefont
  {Linder}}, \bibinfo {author} {\bibfnamefont {Y.}~\bibnamefont {Tanaka}},
  \bibinfo {author} {\bibfnamefont {T.}~\bibnamefont {Yokoyama}}, \bibinfo
  {author} {\bibfnamefont {A.}~\bibnamefont {Sudb\o{}}}, \ and\ \bibinfo
  {author} {\bibfnamefont {N.}~\bibnamefont {Nagaosa}},\ }\href {\doibase
  10.1103/PhysRevLett.104.067001} {\bibfield  {journal} {\bibinfo  {journal}
  {Phys. Rev. Lett.}\ }\textbf {\bibinfo {volume} {104}},\ \bibinfo {pages}
  {067001} (\bibinfo {year} {2010})}\BibitemShut {NoStop}%
\bibitem [{\citenamefont {Sato}\ \emph {et~al.}(2009)\citenamefont {Sato},
  \citenamefont {Takahashi},\ and\ \citenamefont
  {Fujimoto}}]{PhysRevLett.103.020401}%
  \BibitemOpen
  \bibfield  {author} {\bibinfo {author} {\bibfnamefont {M.}~\bibnamefont
  {Sato}}, \bibinfo {author} {\bibfnamefont {Y.}~\bibnamefont {Takahashi}}, \
  and\ \bibinfo {author} {\bibfnamefont {S.}~\bibnamefont {Fujimoto}},\ }\href
  {\doibase 10.1103/PhysRevLett.103.020401} {\bibfield  {journal} {\bibinfo
  {journal} {Phys. Rev. Lett.}\ }\textbf {\bibinfo {volume} {103}},\ \bibinfo
  {pages} {020401} (\bibinfo {year} {2009})}\BibitemShut {NoStop}%
\bibitem [{\citenamefont {Sau}\ \emph {et~al.}(2010{\natexlab{a}})\citenamefont
  {Sau}, \citenamefont {Lutchyn}, \citenamefont {Tewari},\ and\ \citenamefont
  {Das~Sarma}}]{PhysRevLett.104.040502}%
  \BibitemOpen
  \bibfield  {author} {\bibinfo {author} {\bibfnamefont {J.~D.}\ \bibnamefont
  {Sau}}, \bibinfo {author} {\bibfnamefont {R.~M.}\ \bibnamefont {Lutchyn}},
  \bibinfo {author} {\bibfnamefont {S.}~\bibnamefont {Tewari}}, \ and\ \bibinfo
  {author} {\bibfnamefont {S.}~\bibnamefont {Das~Sarma}},\ }\href {\doibase
  10.1103/PhysRevLett.104.040502} {\bibfield  {journal} {\bibinfo  {journal}
  {Phys. Rev. Lett.}\ }\textbf {\bibinfo {volume} {104}},\ \bibinfo {pages}
  {040502} (\bibinfo {year} {2010}{\natexlab{a}})}\BibitemShut {NoStop}%
\bibitem [{\citenamefont {Alicea}(2010)}]{PhysRevB.81.125318}%
  \BibitemOpen
  \bibfield  {author} {\bibinfo {author} {\bibfnamefont {J.}~\bibnamefont
  {Alicea}},\ }\href {\doibase 10.1103/PhysRevB.81.125318} {\bibfield
  {journal} {\bibinfo  {journal} {Phys. Rev. B}\ }\textbf {\bibinfo {volume}
  {81}},\ \bibinfo {pages} {125318} (\bibinfo {year} {2010})}\BibitemShut
  {NoStop}%
\bibitem [{\citenamefont {Oreg}\ \emph {et~al.}(2010)\citenamefont {Oreg},
  \citenamefont {Refael},\ and\ \citenamefont {von
  Oppen}}]{PhysRevLett.105.177002}%
  \BibitemOpen
  \bibfield  {author} {\bibinfo {author} {\bibfnamefont {Y.}~\bibnamefont
  {Oreg}}, \bibinfo {author} {\bibfnamefont {G.}~\bibnamefont {Refael}}, \ and\
  \bibinfo {author} {\bibfnamefont {F.}~\bibnamefont {von Oppen}},\ }\href
  {\doibase 10.1103/PhysRevLett.105.177002} {\bibfield  {journal} {\bibinfo
  {journal} {Phys. Rev. Lett.}\ }\textbf {\bibinfo {volume} {105}},\ \bibinfo
  {pages} {177002} (\bibinfo {year} {2010})}\BibitemShut {NoStop}%
\bibitem [{\citenamefont {Mourik}\ \emph {et~al.}(2012)\citenamefont {Mourik},
  \citenamefont {Zuo}, \citenamefont {Frolov}, \citenamefont {Plissard},
  \citenamefont {Bakkers},\ and\ \citenamefont
  {Kouwenhoven}}]{Science.336.1003}%
  \BibitemOpen
  \bibfield  {author} {\bibinfo {author} {\bibfnamefont {V.}~\bibnamefont
  {Mourik}}, \bibinfo {author} {\bibfnamefont {K.}~\bibnamefont {Zuo}},
  \bibinfo {author} {\bibfnamefont {S.~M.}\ \bibnamefont {Frolov}}, \bibinfo
  {author} {\bibfnamefont {S.~R.}\ \bibnamefont {Plissard}}, \bibinfo {author}
  {\bibfnamefont {E.~P. A.~M.}\ \bibnamefont {Bakkers}}, \ and\ \bibinfo
  {author} {\bibfnamefont {L.~P.}\ \bibnamefont {Kouwenhoven}},\ }\href
  {\doibase 10.1126/science.1222360} {\bibfield  {journal} {\bibinfo  {journal}
  {Science}\ }\textbf {\bibinfo {volume} {336}},\ \bibinfo {pages} {1003}
  (\bibinfo {year} {2012})}\BibitemShut {NoStop}%
\bibitem [{\citenamefont {Wong}\ and\ \citenamefont
  {Law}(2012)}]{arXiv:1211.0338}%
  \BibitemOpen
  \bibfield  {author} {\bibinfo {author} {\bibfnamefont {L.~M.}\ \bibnamefont
  {Wong}}\ and\ \bibinfo {author} {\bibfnamefont {K.~T.}\ \bibnamefont {Law}},\
  }\href {http://arxiv.org/abs/1211.0338} {\bibfield  {journal} {\bibinfo
  {journal} {arXiv:1211.0338}\ } (\bibinfo {year} {2012})}\BibitemShut
  {NoStop}%
\bibitem [{\citenamefont {Sau}\ \emph {et~al.}(2010{\natexlab{b}})\citenamefont
  {Sau}, \citenamefont {Tewari}, \citenamefont {Lutchyn}, \citenamefont
  {Stanescu},\ and\ \citenamefont {Das~Sarma}}]{PhysRevB.82.214509}%
  \BibitemOpen
  \bibfield  {author} {\bibinfo {author} {\bibfnamefont {J.~D.}\ \bibnamefont
  {Sau}}, \bibinfo {author} {\bibfnamefont {S.}~\bibnamefont {Tewari}},
  \bibinfo {author} {\bibfnamefont {R.~M.}\ \bibnamefont {Lutchyn}}, \bibinfo
  {author} {\bibfnamefont {T.~D.}\ \bibnamefont {Stanescu}}, \ and\ \bibinfo
  {author} {\bibfnamefont {S.}~\bibnamefont {Das~Sarma}},\ }\href {\doibase
  10.1103/PhysRevB.82.214509} {\bibfield  {journal} {\bibinfo  {journal} {Phys.
  Rev. B}\ }\textbf {\bibinfo {volume} {82}},\ \bibinfo {pages} {214509}
  (\bibinfo {year} {2010}{\natexlab{b}})}\BibitemShut {NoStop}%
\bibitem [{\citenamefont {Lutchyn}\ \emph {et~al.}(2010)\citenamefont
  {Lutchyn}, \citenamefont {Sau},\ and\ \citenamefont
  {Das~Sarma}}]{PhysRevLett.105.077001}%
  \BibitemOpen
  \bibfield  {author} {\bibinfo {author} {\bibfnamefont {R.~M.}\ \bibnamefont
  {Lutchyn}}, \bibinfo {author} {\bibfnamefont {J.~D.}\ \bibnamefont {Sau}}, \
  and\ \bibinfo {author} {\bibfnamefont {S.}~\bibnamefont {Das~Sarma}},\ }\href
  {\doibase 10.1103/PhysRevLett.105.077001} {\bibfield  {journal} {\bibinfo
  {journal} {Phys. Rev. Lett.}\ }\textbf {\bibinfo {volume} {105}},\ \bibinfo
  {pages} {077001} (\bibinfo {year} {2010})}\BibitemShut {NoStop}%
\bibitem [{\citenamefont {Potter}\ and\ \citenamefont
  {Lee}(2011)}]{PhysRevB.83.094525}%
  \BibitemOpen
  \bibfield  {author} {\bibinfo {author} {\bibfnamefont {A.~C.}\ \bibnamefont
  {Potter}}\ and\ \bibinfo {author} {\bibfnamefont {P.~A.}\ \bibnamefont
  {Lee}},\ }\href {\doibase 10.1103/PhysRevB.83.094525} {\bibfield  {journal}
  {\bibinfo  {journal} {Phys. Rev. B}\ }\textbf {\bibinfo {volume} {83}},\
  \bibinfo {pages} {094525} (\bibinfo {year} {2011})}\BibitemShut {NoStop}%
\bibitem [{\citenamefont {Tewari}\ \emph {et~al.}(2011)\citenamefont {Tewari},
  \citenamefont {Stanescu}, \citenamefont {Sau},\ and\ \citenamefont
  {Sarma}}]{NewJPhys.13.065004}%
  \BibitemOpen
  \bibfield  {author} {\bibinfo {author} {\bibfnamefont {S.}~\bibnamefont
  {Tewari}}, \bibinfo {author} {\bibfnamefont {T.~D.}\ \bibnamefont
  {Stanescu}}, \bibinfo {author} {\bibfnamefont {J.~D.}\ \bibnamefont {Sau}}, \
  and\ \bibinfo {author} {\bibfnamefont {S.~D.}\ \bibnamefont {Sarma}},\ }\href
  {http://stacks.iop.org/1367-2630/13/i=6/a=065004} {\bibfield  {journal}
  {\bibinfo  {journal} {New J. Phys.}\ }\textbf {\bibinfo {volume} {13}},\
  \bibinfo {pages} {065004} (\bibinfo {year} {2011})}\BibitemShut {NoStop}%
\bibitem [{\citenamefont {Stanescu}\ \emph {et~al.}(2011)\citenamefont
  {Stanescu}, \citenamefont {Lutchyn},\ and\ \citenamefont
  {Das~Sarma}}]{PhysRevB.84.144522}%
  \BibitemOpen
  \bibfield  {author} {\bibinfo {author} {\bibfnamefont {T.~D.}\ \bibnamefont
  {Stanescu}}, \bibinfo {author} {\bibfnamefont {R.~M.}\ \bibnamefont
  {Lutchyn}}, \ and\ \bibinfo {author} {\bibfnamefont {S.}~\bibnamefont
  {Das~Sarma}},\ }\href {\doibase 10.1103/PhysRevB.84.144522} {\bibfield
  {journal} {\bibinfo  {journal} {Phys. Rev. B}\ }\textbf {\bibinfo {volume}
  {84}},\ \bibinfo {pages} {144522} (\bibinfo {year} {2011})}\BibitemShut
  {NoStop}%
\bibitem [{\citenamefont {Lutchyn}\ \emph {et~al.}(2011)\citenamefont
  {Lutchyn}, \citenamefont {Stanescu},\ and\ \citenamefont
  {Das~Sarma}}]{PhysRevLett.106.127001}%
  \BibitemOpen
  \bibfield  {author} {\bibinfo {author} {\bibfnamefont {R.~M.}\ \bibnamefont
  {Lutchyn}}, \bibinfo {author} {\bibfnamefont {T.~D.}\ \bibnamefont
  {Stanescu}}, \ and\ \bibinfo {author} {\bibfnamefont {S.}~\bibnamefont
  {Das~Sarma}},\ }\href {\doibase 10.1103/PhysRevLett.106.127001} {\bibfield
  {journal} {\bibinfo  {journal} {Phys. Rev. Lett.}\ }\textbf {\bibinfo
  {volume} {106}},\ \bibinfo {pages} {127001} (\bibinfo {year}
  {2011})}\BibitemShut {NoStop}%
\bibitem [{\citenamefont {Kim}\ \emph {et~al.}(2012)\citenamefont {Kim},
  \citenamefont {Cano},\ and\ \citenamefont {Nayak}}]{arXiv:1208.3701}%
  \BibitemOpen
  \bibfield  {author} {\bibinfo {author} {\bibfnamefont {Y.}~\bibnamefont
  {Kim}}, \bibinfo {author} {\bibfnamefont {J.}~\bibnamefont {Cano}}, \ and\
  \bibinfo {author} {\bibfnamefont {C.}~\bibnamefont {Nayak}},\ }\href
  {http://arxiv.org/abs/1208.3701} {\bibfield  {journal} {\bibinfo  {journal}
  {arXiv:1208.3701}\ } (\bibinfo {year} {2012})}\BibitemShut {NoStop}%
\bibitem [{\citenamefont {Schnyder}\ \emph {et~al.}(2008)\citenamefont
  {Schnyder}, \citenamefont {Ryu}, \citenamefont {Furusaki},\ and\
  \citenamefont {Ludwig}}]{PhysRevB.78.195125}%
  \BibitemOpen
  \bibfield  {author} {\bibinfo {author} {\bibfnamefont {A.~P.}\ \bibnamefont
  {Schnyder}}, \bibinfo {author} {\bibfnamefont {S.}~\bibnamefont {Ryu}},
  \bibinfo {author} {\bibfnamefont {A.}~\bibnamefont {Furusaki}}, \ and\
  \bibinfo {author} {\bibfnamefont {A.~W.~W.}\ \bibnamefont {Ludwig}},\ }\href
  {\doibase 10.1103/PhysRevB.78.195125} {\bibfield  {journal} {\bibinfo
  {journal} {Phys. Rev. B}\ }\textbf {\bibinfo {volume} {78}},\ \bibinfo
  {pages} {195125} (\bibinfo {year} {2008})}\BibitemShut {NoStop}%
\bibitem [{\citenamefont {Ryu}\ \emph {et~al.}(2010)\citenamefont {Ryu},
  \citenamefont {Schnyder}, \citenamefont {Furusaki},\ and\ \citenamefont
  {Ludwig}}]{NewJPhys.12.065010}%
  \BibitemOpen
  \bibfield  {author} {\bibinfo {author} {\bibfnamefont {S.}~\bibnamefont
  {Ryu}}, \bibinfo {author} {\bibfnamefont {A.~P.}\ \bibnamefont {Schnyder}},
  \bibinfo {author} {\bibfnamefont {A.}~\bibnamefont {Furusaki}}, \ and\
  \bibinfo {author} {\bibfnamefont {A.~W.~W.}\ \bibnamefont {Ludwig}},\ }\href
  {\doibase 10.1088/1367-2630/12/6/065010} {\bibfield  {journal} {\bibinfo
  {journal} {New J. Phys.}\ }\textbf {\bibinfo {volume} {12}},\ \bibinfo
  {pages} {065010} (\bibinfo {year} {2010})}\BibitemShut {NoStop}%
\bibitem [{\citenamefont {Mackenzie}\ and\ \citenamefont
  {Maeno}(2003)}]{RevModPhys.75.657}%
  \BibitemOpen
  \bibfield  {author} {\bibinfo {author} {\bibfnamefont {A.~P.}\ \bibnamefont
  {Mackenzie}}\ and\ \bibinfo {author} {\bibfnamefont {Y.}~\bibnamefont
  {Maeno}},\ }\href {\doibase 10.1103/RevModPhys.75.657} {\bibfield  {journal}
  {\bibinfo  {journal} {Rev. Mod. Phys.}\ }\textbf {\bibinfo {volume} {75}},\
  \bibinfo {pages} {657} (\bibinfo {year} {2003})}\BibitemShut {NoStop}%
\bibitem [{\citenamefont {Tanaka}\ \emph {et~al.}(2009)\citenamefont {Tanaka},
  \citenamefont {Yokoyama}, \citenamefont {Balatsky},\ and\ \citenamefont
  {Nagaosa}}]{PhysRevB.79.060505}%
  \BibitemOpen
  \bibfield  {author} {\bibinfo {author} {\bibfnamefont {Y.}~\bibnamefont
  {Tanaka}}, \bibinfo {author} {\bibfnamefont {T.}~\bibnamefont {Yokoyama}},
  \bibinfo {author} {\bibfnamefont {A.~V.}\ \bibnamefont {Balatsky}}, \ and\
  \bibinfo {author} {\bibfnamefont {N.}~\bibnamefont {Nagaosa}},\ }\href
  {\doibase 10.1103/PhysRevB.79.060505} {\bibfield  {journal} {\bibinfo
  {journal} {Phys. Rev. B}\ }\textbf {\bibinfo {volume} {79}},\ \bibinfo
  {pages} {060505} (\bibinfo {year} {2009})}\BibitemShut {NoStop}%
\bibitem [{\citenamefont {Sato}\ and\ \citenamefont
  {Fujimoto}(2009)}]{PhysRevB.79.094504}%
  \BibitemOpen
  \bibfield  {author} {\bibinfo {author} {\bibfnamefont {M.}~\bibnamefont
  {Sato}}\ and\ \bibinfo {author} {\bibfnamefont {S.}~\bibnamefont
  {Fujimoto}},\ }\href {\doibase 10.1103/PhysRevB.79.094504} {\bibfield
  {journal} {\bibinfo  {journal} {Phys. Rev. B}\ }\textbf {\bibinfo {volume}
  {79}},\ \bibinfo {pages} {094504} (\bibinfo {year} {2009})}\BibitemShut
  {NoStop}%
\bibitem [{\citenamefont {Nakosai}\ \emph {et~al.}(2012)\citenamefont
  {Nakosai}, \citenamefont {Tanaka},\ and\ \citenamefont
  {Nagaosa}}]{PhysRevLett.108.147003}%
  \BibitemOpen
  \bibfield  {author} {\bibinfo {author} {\bibfnamefont {S.}~\bibnamefont
  {Nakosai}}, \bibinfo {author} {\bibfnamefont {Y.}~\bibnamefont {Tanaka}}, \
  and\ \bibinfo {author} {\bibfnamefont {N.}~\bibnamefont {Nagaosa}},\ }\href
  {\doibase 10.1103/PhysRevLett.108.147003} {\bibfield  {journal} {\bibinfo
  {journal} {Phys. Rev. Lett.}\ }\textbf {\bibinfo {volume} {108}},\ \bibinfo
  {pages} {147003} (\bibinfo {year} {2012})}\BibitemShut {NoStop}%
\bibitem [{\citenamefont {Wang}\ and\ \citenamefont
  {Zhang}(2012)}]{PhysRevX.2.031008}%
  \BibitemOpen
  \bibfield  {author} {\bibinfo {author} {\bibfnamefont {Z.}~\bibnamefont
  {Wang}}\ and\ \bibinfo {author} {\bibfnamefont {S.-C.}\ \bibnamefont
  {Zhang}},\ }\href {\doibase 10.1103/PhysRevX.2.031008} {\bibfield  {journal}
  {\bibinfo  {journal} {Phys. Rev. X}\ }\textbf {\bibinfo {volume} {2}},\
  \bibinfo {pages} {031008} (\bibinfo {year} {2012})}\BibitemShut {NoStop}%
\bibitem [{\citenamefont {Budich}\ and\ \citenamefont
  {Trauzettel}(2012)}]{arXiv:1207.1104}%
  \BibitemOpen
  \bibfield  {author} {\bibinfo {author} {\bibfnamefont {J.~C.}\ \bibnamefont
  {Budich}}\ and\ \bibinfo {author} {\bibfnamefont {B.}~\bibnamefont
  {Trauzettel}},\ }\href {http://arxiv.org/abs/1207.1104} {\bibfield  {journal}
  {\bibinfo  {journal} {arXiv:1207.1104}\ } (\bibinfo {year}
  {2012})}\BibitemShut {NoStop}%
\bibitem [{\citenamefont {Kitaev}(2001)}]{PhysUsp.44.131}%
  \BibitemOpen
  \bibfield  {author} {\bibinfo {author} {\bibfnamefont {A.~Y.}\ \bibnamefont
  {Kitaev}},\ }\href {http://stacks.iop.org/1063-7869/44/i=10S/a=S29}
  {\bibfield  {journal} {\bibinfo  {journal} {Phys.-Usp.}\ }\textbf {\bibinfo
  {volume} {44}},\ \bibinfo {pages} {131} (\bibinfo {year} {2001})}\BibitemShut
  {NoStop}%
\bibitem [{\citenamefont {Sau}\ \emph {et~al.}(2011{\natexlab{a}})\citenamefont
  {Sau}, \citenamefont {Clarke},\ and\ \citenamefont
  {Tewari}}]{PhysRevB.84.094505}%
  \BibitemOpen
  \bibfield  {author} {\bibinfo {author} {\bibfnamefont {J.~D.}\ \bibnamefont
  {Sau}}, \bibinfo {author} {\bibfnamefont {D.~J.}\ \bibnamefont {Clarke}}, \
  and\ \bibinfo {author} {\bibfnamefont {S.}~\bibnamefont {Tewari}},\ }\href
  {\doibase 10.1103/PhysRevB.84.094505} {\bibfield  {journal} {\bibinfo
  {journal} {Phys. Rev. B}\ }\textbf {\bibinfo {volume} {84}},\ \bibinfo
  {pages} {094505} (\bibinfo {year} {2011}{\natexlab{a}})}\BibitemShut
  {NoStop}%
\bibitem [{\citenamefont {Sau}\ \emph {et~al.}(2011{\natexlab{b}})\citenamefont
  {Sau}, \citenamefont {Halperin}, \citenamefont {Flensberg},\ and\
  \citenamefont {Das~Sarma}}]{PhysRevB.84.144509}%
  \BibitemOpen
  \bibfield  {author} {\bibinfo {author} {\bibfnamefont {J.~D.}\ \bibnamefont
  {Sau}}, \bibinfo {author} {\bibfnamefont {B.~I.}\ \bibnamefont {Halperin}},
  \bibinfo {author} {\bibfnamefont {K.}~\bibnamefont {Flensberg}}, \ and\
  \bibinfo {author} {\bibfnamefont {S.}~\bibnamefont {Das~Sarma}},\ }\href
  {\doibase 10.1103/PhysRevB.84.144509} {\bibfield  {journal} {\bibinfo
  {journal} {Phys. Rev. B}\ }\textbf {\bibinfo {volume} {84}},\ \bibinfo
  {pages} {144509} (\bibinfo {year} {2011}{\natexlab{b}})}\BibitemShut
  {NoStop}%
\bibitem [{\citenamefont {Halperin}\ \emph {et~al.}(2012)\citenamefont
  {Halperin}, \citenamefont {Oreg}, \citenamefont {Stern}, \citenamefont
  {Refael}, \citenamefont {Alicea},\ and\ \citenamefont {von
  Oppen}}]{PhysRevB.85.144501}%
  \BibitemOpen
  \bibfield  {author} {\bibinfo {author} {\bibfnamefont {B.~I.}\ \bibnamefont
  {Halperin}}, \bibinfo {author} {\bibfnamefont {Y.}~\bibnamefont {Oreg}},
  \bibinfo {author} {\bibfnamefont {A.}~\bibnamefont {Stern}}, \bibinfo
  {author} {\bibfnamefont {G.}~\bibnamefont {Refael}}, \bibinfo {author}
  {\bibfnamefont {J.}~\bibnamefont {Alicea}}, \ and\ \bibinfo {author}
  {\bibfnamefont {F.}~\bibnamefont {von Oppen}},\ }\href {\doibase
  10.1103/PhysRevB.85.144501} {\bibfield  {journal} {\bibinfo  {journal} {Phys.
  Rev. B}\ }\textbf {\bibinfo {volume} {85}},\ \bibinfo {pages} {144501}
  (\bibinfo {year} {2012})}\BibitemShut {NoStop}%
\bibitem [{\citenamefont {Fu}(2010)}]{PhysRevLett.104.056402}%
  \BibitemOpen
  \bibfield  {author} {\bibinfo {author} {\bibfnamefont {L.}~\bibnamefont
  {Fu}},\ }\href {\doibase 10.1103/PhysRevLett.104.056402} {\bibfield
  {journal} {\bibinfo  {journal} {Phys. Rev. Lett.}\ }\textbf {\bibinfo
  {volume} {104}},\ \bibinfo {pages} {056402} (\bibinfo {year}
  {2010})}\BibitemShut {NoStop}%
\bibitem [{\citenamefont {Tada}\ \emph {et~al.}(2009)\citenamefont {Tada},
  \citenamefont {Kawakami},\ and\ \citenamefont
  {Fujimoto}}]{NewJPhys.11.055070}%
  \BibitemOpen
  \bibfield  {author} {\bibinfo {author} {\bibfnamefont {Y.}~\bibnamefont
  {Tada}}, \bibinfo {author} {\bibfnamefont {N.}~\bibnamefont {Kawakami}}, \
  and\ \bibinfo {author} {\bibfnamefont {S.}~\bibnamefont {Fujimoto}},\ }\href
  {http://stacks.iop.org/1367-2630/11/i=5/a=055070} {\bibfield  {journal}
  {\bibinfo  {journal} {New J. Phys.}\ }\textbf {\bibinfo {volume} {11}},\
  \bibinfo {pages} {055070} (\bibinfo {year} {2009})}\BibitemShut {NoStop}%
\bibitem [{\citenamefont {Mizukami}\ \emph {et~al.}(2011)\citenamefont
  {Mizukami}, \citenamefont {Shishido}, \citenamefont {Shibauchi},
  \citenamefont {Shimozawa}, \citenamefont {Yasumoto}, \citenamefont
  {Watanabe}, \citenamefont {Yamashita}, \citenamefont {Ikeda}, \citenamefont
  {Terashima}, \citenamefont {Kontani},\ and\ \citenamefont
  {Matsuda}}]{NatPhys.7.849}%
  \BibitemOpen
  \bibfield  {author} {\bibinfo {author} {\bibfnamefont {Y.}~\bibnamefont
  {Mizukami}}, \bibinfo {author} {\bibfnamefont {H.}~\bibnamefont {Shishido}},
  \bibinfo {author} {\bibfnamefont {T.}~\bibnamefont {Shibauchi}}, \bibinfo
  {author} {\bibfnamefont {M.}~\bibnamefont {Shimozawa}}, \bibinfo {author}
  {\bibfnamefont {S.}~\bibnamefont {Yasumoto}}, \bibinfo {author}
  {\bibfnamefont {D.}~\bibnamefont {Watanabe}}, \bibinfo {author}
  {\bibfnamefont {M.}~\bibnamefont {Yamashita}}, \bibinfo {author}
  {\bibfnamefont {H.}~\bibnamefont {Ikeda}}, \bibinfo {author} {\bibfnamefont
  {T.}~\bibnamefont {Terashima}}, \bibinfo {author} {\bibfnamefont
  {H.}~\bibnamefont {Kontani}}, \ and\ \bibinfo {author} {\bibfnamefont
  {Y.}~\bibnamefont {Matsuda}},\ }\href {\doibase 10.1038/nphys2112} {\bibfield
   {journal} {\bibinfo  {journal} {Nature Phys.}\ }\textbf {\bibinfo {volume}
  {7}},\ \bibinfo {pages} {849} (\bibinfo {year} {2011})}\BibitemShut {NoStop}%
\end{thebibliography}
\end{document}